\documentclass[]{revtex4}
\usepackage{amsfonts}
\usepackage{amssymb,epsf}
\usepackage{graphicx}
\begin{document}
\title{Quantum corrections to  thermodynamics of quasitopological  black holes  }
\author{Sudhaker Upadhyay \footnote{Presently at K.L.S. College Nawada, Nawada-805110, India}
 }
 \email{sudhakerupadhyay@gmail.com}
 \affiliation {Centre for Theoretical Studies, Indian Institute of Technology Kharagpur,  Kharagpur-721302, WB, India}

\begin{abstract}Based on the modification to area-law due to thermal fluctuation at small horizon 
radius, we investigate the 
thermodynamics of charged quasitopological and charged rotating quasitopological black holes.
In particular, we derive the leading-order corrections to the Gibbs free energy, charge and 
total mass densities. In order to analyse the behavior  of the thermal fluctuations on the thermodynamics of small black holes,  we draw a comparative analysis  between the first-order  corrected and original thermodynamical quantities. We also examine the stability and bound points of such black holes under  effect of leading-order corrections. 
\end{abstract}

\maketitle
 
\section{Overview and motivation}
According to AdS/CFT duality, the Einstein general relativity in the
bulk space-time corresponds to a gauge theory  living on
the boundary  with a large
$N$ (number of colors) and large 't Hooft coupling \cite{Mal}.  
 Since the coupling constants in
the gravity side relate to central charges in the gauge theory, 
therefore   Einstein gravity  has limited number of dual CFTs, in particular  only those CFTs
which have equal  central charges,  as Einstein gravity does not have enough free parameters.  The presence of  various higher-order derivatives in AdS gravity
corresponds to new couplings among operators in the dual CFT.
One well-know example of higher derivatives gravity theories  is Gauss-Bonnet gravity. The Gauss-Bonnet gravity involves  only
one quadratic coupling term and therefore   the corresponding range of dual theory is still limited.
In order to improve this limitation of  holographic studies to the classes of CFTs,  one
  has to introduce  
 the new  higher order curvature terms, atleast  a curvature-cubed terms,   into gravity.
One may achieve such a curvature-cubed interactions by adding the cubic term in Lovelock gravity \cite{hof}, but can not be very helpful as such
term is topological in nature  and becomes significant  only in very high dimensions.

 Recently,  a new toy model for gravitation
action has been proposed which contains not only the Gauss-Bonnet term but also a
 curvature-cubed interaction \cite{may}.  This is a quasitopological
gravity model as the cubic terms  do  not have   a topological origin like Lovelock gravity but contribute dynamically  to the evolution of fields in the bulk. 
   This quasitopological gravity theory is endowed with two important properties. First,  the equations of motion are only second order in derivatives, and second there exists the exact black hole solutions \cite{may}.
 The holographic discussions for these black
hole solutions with some recipes of AdS/CFT  duality have been given in 
 \cite{may1}. Recently,  the surface term of quasitopological
gravity for space-time  with flat boundary is introduced and 
   the thermodynamic
properties of these solutions have been investigated by using the relation
between on-shell action and Gibbs free energy \cite{l1}.

An important discovery that black holes behave as thermodynamic objects had
affected  our understanding of gravity theory and its relationship to quantum field theory considerably. Bekenstein  and Hawking  were first who proposed that black holes radiate as black bodies with characteristic   entropy related to the area of the horizon \cite{bak}.  
In present scenario, it is more or less certain that black holes much larger than the Planck scale have entropy proportional to its horizon area \cite{bak,str,ast,car,sol}.
So, this poses an interesting question  that   what could be the leading-order corrections when one reduces the size of the  black holes. 
To answer this question, several attempts have been made. 
For instance,  using a corrected version
of the asymptotic Cardy formula  for    BTZ, string theoretic and all other black holes, whose
microscopic degrees of freedom are described by an underlying CFT  \cite{car1},  the leading-order  corrections have found logarithmic in nature. In fact, the consideration of matter fields in black hole backgrounds also yields logarithmic correction to the  black holes  entropy at  the leading order    \cite{man}. The leading-order correction to black holes entropy is also logarithmic  by considering string-black hole
correspondence \cite{sol1} and   using Rademacher expansion of the partition function \cite{bir}.  Furthermore,   Das et al. in Ref. \cite{das} showed that the
 leading-order corrections to the entropy of any thermodynamic system due to small
statistical fluctuations around equilibrium are always logarithmic.

The study  of leading-order correction to the black holes thermodynamics is a subject 
of current interests. In this direction, recently,  the  effects of quantum corrections on thermodynamics and stability  of  G\"{o}del black hole \cite{godel},
 Schwarzschild-Beltrami-de Sitter black hole \cite{sud2} and   massive black hole in  AdS space
 \cite{sud3} have been studied.  The  corrected thermodynamics of a  dilatonic black hole has also been discussed \cite{jy} which meets the same universal form of correction term. In another work, the corrected thermodynamics
of a black hole is also studied from  the partition function points of view \cite{bss}. 
 The quantum gravity effects on the Ho\v rava-Lifshitz black hole thermodynamics are   analysed 
 and their stability  is also discussed  \cite{sud1}.  Similar investigation in case of the modified Hayward black hole is also made, where it has been found that correction term reduces the pressure and internal energy
of the Hayward black hole \cite{behn}. We try to extend such study to the case of 
quasitopological black holes.

In this paper,  we consider a charged quasitopological model which exhibits black hole solutions 
and discuss the effects of leading-order correction on thermodynamics which becomes significant for small size of the black holes.  First, we compute the  leading-order correction to the 
entropy of charged quasitopological black hole and plot a graph to make
a comparative analysis between corrected and  uncorrected entropy densities for smaller black holes. Here, we find that for (negative-)positive correction parameter  ($\alpha$)  there exists
a (positive-)negative peak for the  corrected entropy density at sufficiently small black holes.  The corrected   entropy density  
becomes negative valued for the positive correction parameter, which is not physical and therefore can be forbidden.   We see that  two   critical points    exist for the entropy density. The correction term affects significantly the  entropy densities in between these critical points. Furthermore, we derive the first-order corrected Gibbs free energy density and discuss the effects of correction terms.
 We observe that the correction terms 
with negative correction parameter make  Gibbs free energy density (more-)less negative valued for the (smaller-)larger black holes.
However, the correction terms with positive correction parameter   make  Gibbs free energy density more positive valued  for the black holes with  smaller horizon radius.  For the larger values of charge and AdS radius, the deviation of   corrected Gibbs free energy density  with their original value becomes less. We also calculated the corrected expression for the total charge of the
quasitopological black holes which coincides with their original expression in limit $\alpha\rightarrow 0$. Moreover, we evaluate the first-order corrected expression for 
the mass density of this black hole. We find that a critical point exists  for total mass density below which corrected terms with the positive correction parameter shows opposite behavior. 
We also check the stability and  bound point of black holes by calculating  specific heat at constant chemical potential and plot 
with respect to horizon radius. We find that the phase transition does not occur due to the correction term  with positive correction parameter and black holes are in stable state. The   correction term with negative parameter causes instability for such black holes. Furthermore, in the same fashion, we investigate the  effects of thermal fluctuation on the  thermodynamics of charged quasitopological black holes endowed with global rotation.

The paper is organized as following. In section \ref{2}, we 
derive the corrected expression for entropy density due to the thermal fluctuations when   the size of the black holes is reduced  to  the Planck scale. In section \ref{3}, 
we discuss the effects of quantum corrections due to thermal fluctuations on the thermodynamics of charged quasitopological black holes. Within this section, we study the 
influence of leading-order correction  on stability of such black holes. 
In section \ref{4}, we consider a charged topological black holes endowed with global rotation
and discuss  the effects of thermal fluctuations on the thermodynamics of it. We also study the  stability and bound points of charged rotating quasitopological black holes under the influence of 
thermal fluctuations.  We summarize our results with concluding remarks in the last section \ref{5}.

\section{Thermodynamics under (quantum)  thermal instability: Preliminaries}\label{2}

  In this section, we review the  corrections to thermodynamic entropy density
of the quasitopological  black holes when small stable fluctuations around equilibrium are taken into account.  In this connection, one may assume that the system of quasitopological  black holes is characterized by the 
  canonical ensemble.   In order to begin the analysis,   let us  first
define the density of states with fixed energy  as \cite{boh,rk}
\begin{eqnarray}
\rho(E) =\frac{1}{2\pi i}\int_{c-i\infty}^{c+i\infty}e^{{\cal S}(\beta)}d\beta.\label{rho}
\end{eqnarray}
Here ${\cal S}(\beta)$ refers to  the exact entropy  density which is not just its value at equilibrium and depends on temperature $T=1/\beta$ explicitly. The exact entropy  density corresponds to the
 sum of entropy densities of subsystems of the thermodynamical system, which are small enough to be considered in equilibrium. 
In order to solve the  complex integral (\ref{rho}), we utilize the method of steepest descent around the saddle point $\beta_0 (={1/T_H})$ such that $\left(\frac{\partial {\cal S}(\beta)}{\partial \beta}\right)_{\beta=\beta_0}=0$. We assume  that the quasitopological black hole is in equilibrium at Hawking temperature $T_H$.
Now, the Taylor expansion of  exact entropy  density around the saddle point $\beta=\beta_0$ 
yields
\begin{eqnarray}
{\cal S}(\beta)=S_0+\frac{1}{2}(\beta-\beta_0)^2 \left(\frac{\partial^2 {\cal S}(\beta)}{\partial \beta^2}\right)_{\beta=\beta_0}+ \mbox{(higher order terms)},\label{s}
\end{eqnarray}
where $S_0={\cal S}(\beta_0)$ refers the leading-order entropy  density.
Now, by plugging this ${\cal S}(\beta)$   (\ref{s}) into (\ref{rho}), and solving integral
by choosing $c=\beta_0$ for positive $\left(\frac{\partial^2 {\cal S}(\beta)}{\partial \beta^2}\right)_{\beta=\beta_0}$ 
leads to \cite{das}
\begin{eqnarray}
\rho(E)=\frac{e^{S_0}}{\sqrt{2\pi \left(\frac{\partial^2 {\cal S}(\beta)}{\partial \beta^2}\right)_{\beta=\beta_0}}}.
\end{eqnarray}
 The logarithm of the above density of states yields
 the corrected microcanonical entropy  density at equilibrium (obtained by incorporating small fluctuations around
thermal equilibrium)
\begin{eqnarray}
S=S_0-\frac{1}{2}\log \left(\frac{\partial^2 {\cal S}(\beta)}{\partial \beta^2}\right)_{\beta=\beta_0}+ \mbox{(sub-leading terms)}.
\end{eqnarray}
 By considering the most general form  of the exact entropy  density, ${\cal S}(\beta)$, the form of   $\left(\frac{\partial^2 {\cal S}(\beta)}{\partial \beta^2}\right)_{\beta=\beta_0}$ can be determined.
The  generic expression for leading-order correction  to Bekenstein-Hawking formula is 
calculated by \cite{das}
\begin{equation}\label{correctedS}
S = S_{0} + \alpha \ln (S_{0} T_H^{2}),
\end{equation}
where $\alpha$ is  a (constant) correction parameter.  One should note that  we considered a general correction parameter $\alpha$ because this is not fixed valued and  takes different values in different circumstances. Eventually, we observe that the leading-order corrections to the entropy   density of any thermodynamic system (quasitopological  black holes)   due to small
statistical fluctuations around equilibrium are logarithmic in nature. Now, we shall study the
effects of such correction term on the thermodynamics of both the charged and charged rotating quasitopological  black holes. 
\section{Charged quasitopological black holes: Thermal instability} \label{3}
The general action for the quasitopological gravity with  cosmological constant $\Lambda$ in $(d + 1)$ space-time dimensions in the presence of the electromagnetic field ($A_b$)
can be given as \cite{may, l1}
\begin{eqnarray}
I&=&\frac{1}{16\pi G_{d+1}}\int d^{d+1}x \sqrt{-g}\left[R-\Lambda+\frac{\lambda l^2}{(d-2)(d-3)}
\mathcal{X}_2 \right.\nonumber\\
&+&\left.\frac{8(2d-1)\mu l^4}{(d-2)(d-5)(3d^2-9d+4)}\mathcal{X}_3 -\frac{1}{4}F_{ab}F^{ab} \right],\label{act}
\end{eqnarray}
where $\Lambda =- {d(d-1)}/{2l^2}$, Maxwell field-strength tensor $F_{ab}=\partial_a A_b -\partial_b A_a$, $\lambda$ is the Gauss-Bonnet coupling constant and $\mu$ is the quasitopological coupling constant. Here,   $\mathcal{X}_2$  and  $\mathcal{X}_3$   are the
Gauss-Bonnet   and  quasitopological terms, respectively, with  following explicit expressions:
\begin{eqnarray}
\mathcal{X}_2 &=&R_{abcd}R^{abcd}-4\,R_{ab}R^{ab}+R^2,\\
\mathcal{X}_3 &=& R_{a\,\,b}^{\,\,c\,\,\,d} R_{c\,\,d}^{\,\,e\,\,\,f}
R_{e\,\,f}^{\,\,a\,\,\,b} +
\frac{1}{(2d-1)(d-3)}\left[\frac{3(3d-5)}{8}R_{a b c d}R^{a b c d}
R \right.\nonumber\\
&-&  3(d-1) R_{a b c d}R^{a b c}{}_{e}R^{d e}+ 3(d+1)R_{a b c d} R^{a c}R^{b d} \\
&+& \left.\,6(d-1)R_a{}^{b}R_b{}^{c}R_c{}^{a}-\frac{3(3d-1)}{2} R_a^{\,\,b}R_b^{\,\,a}R
+\frac{3(d+1)}{8}R^3\right]. 
\end{eqnarray}
Now, in order to study the thermodynamics of quasitopological black hole described by the action (\ref{act}),
 we consider a $(d+1)$-dimensional static metric with a flat boundary as follows,
\begin{eqnarray}
ds^2=N^2(r)f(r)dt^2+\frac{dr^2}{f(r)}+r^2\sum_{i=1}^{d-1}d\phi_i^2,
\end{eqnarray}
where $N(r)$ is a lapse function.
Here we should note that the field equations of quasitopological gravity are second-order differential equations  only  for this metric. 
It has been shown in Ref.  \cite{l1} that $N(r)$ must be a constant and therefore   without loss of generality can be   set to
unit (i.e. $N(r)=1$).
By considering the gauge potential ansatz $A_a =h(r)\delta^t_a$ to have radial electric field,
the equation of motion will have the following solution: $h(r)=-\sqrt{\frac{2(n-1)}{ n-2 }}\frac{q}{r^{d-2}}$. Now, the solution for metric function  is given by \cite{l1}
\begin{eqnarray}
f(r)=\frac{r^2}{l^2}-\frac{m}{r^{d-2}}+\frac{q^2}{r^{2(d-2)}},\label{f}
\end{eqnarray}
where $q$ and  $m$ are integration constants, respectively,
related to the electric charge   and total mass of the quasitopological black holes.
The integration constant $m$ can easily be  evaluated from the metric function on the horizon ($f (r = r+) = 0$) as 
\begin{eqnarray}
m=\frac{r_+^d}{l^2}+\frac{q^2}{r_+^{d-2}}.\label{m}
\end{eqnarray}
Now, exploiting relations (\ref{f}) and (\ref{m}), the  Hawking  temperature of the event horizon  can be calculated by \cite{l1}
\begin{eqnarray}
T_H=\left.\frac{f'(r)}{4\pi}\right|_{r=r_+}=\frac{dr_+-(d-2)q^2l^2r^{3-2d}}{4\pi l^2},\label{t}
\end{eqnarray}
where $r_+$ is the outer horizon radius. The  
Gibbs free energy   per
unit volume can be identified 
with the Euclidean action per volume times the temperature \cite{gib}. 
Corresponding to the resulting Gibbs free energy   per
unit volume, the leading entropy density of charged quasitopological black holes is calculated as \cite{l1}
\begin{eqnarray}
S_0=\frac{1}{4}r_+^{d-1}.\label{ss}
\end{eqnarray}
The expression for electric potential, measured at infinity with respect
to the horizon,  for static case is given by \cite{l1,cve}
\begin{eqnarray}
\Phi=\sqrt{\frac{2(d-1)}{d-2}}\frac{q}{r_+^{d-2}}.\label{phi}
\end{eqnarray}
With the help of relations (\ref{t}) and (\ref{phi}), the  Hawking  temperature of quasitopological  black holes can be expressed in terms of  electric potential as following:
\begin{eqnarray}
T_H= \frac{dr_+}{4\pi l^2}-\frac{(d-2)^2\Phi^2}{8\pi (d-1) r_+},
\end{eqnarray}
which leads to the horizon radius  in terms of temperature and electric potential  as
\begin{eqnarray}
r_+=\frac{2\pi l^2}{d}T+2\left[\frac{\pi^2 l^4}{d^2}T^2+\frac{(d-2)^2 l^2}{8d(d-1)}\Phi^2
\right]^{1/2}.
\end{eqnarray}
Utilizing the relations (\ref{t}) and (\ref{ss}), the first-order corrected entropy per volume (\ref{correctedS}) for the charged quasitopological black hole due to the thermal fluctuation
 is computed as
\begin{eqnarray}
S&=&\frac{1}{4}r_+^{d-1}+\alpha \log\left[\frac{d^2r_+^{d+1}+(d-2)^2q^4l^4 r_+^{5-3d}-2d(d-2)q^2l^2r_+^{3-d}}{64\pi^2 l^4}\right].
\end{eqnarray}
This can further be expressed  in terms of electric potential as following:
 \begin{eqnarray}
S=\frac{1}{4}r_+^{d-1}+\alpha \log\left[\frac{1}{64\pi^2}\left(\frac{d^2r^{d+1}}{l^4}+\frac{(d-2)^4}{4(d-1)^2}\Phi^4r_+^{d-3}-\frac{d(d-2)^2}{(d-1)}\frac{\Phi^2r_+^{d-1}}{l^2}\right) \right].
\end{eqnarray}

 \begin{figure}[h!]
 \begin{center}$
 \begin{array}{cc}
\includegraphics[width=70 mm]{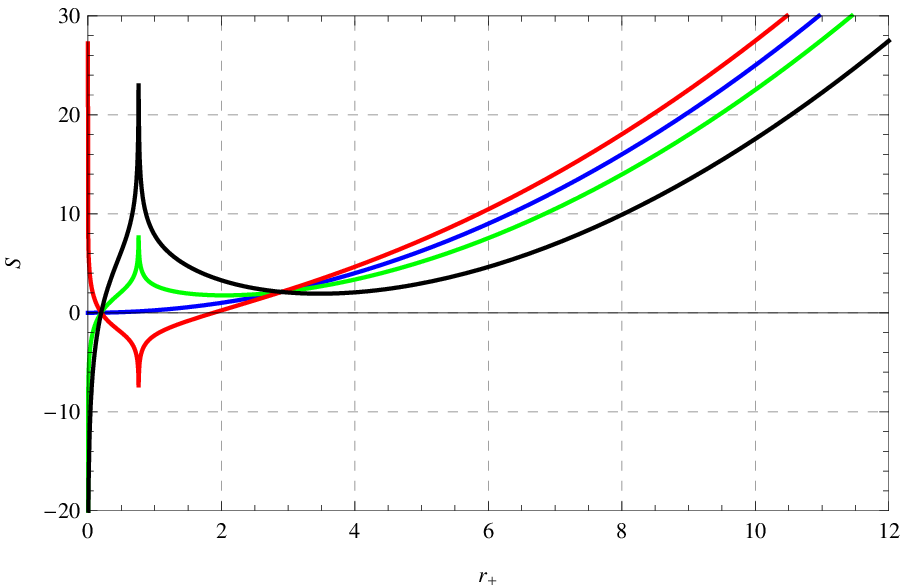}  \ \ \ \ & \includegraphics[width=70 mm]{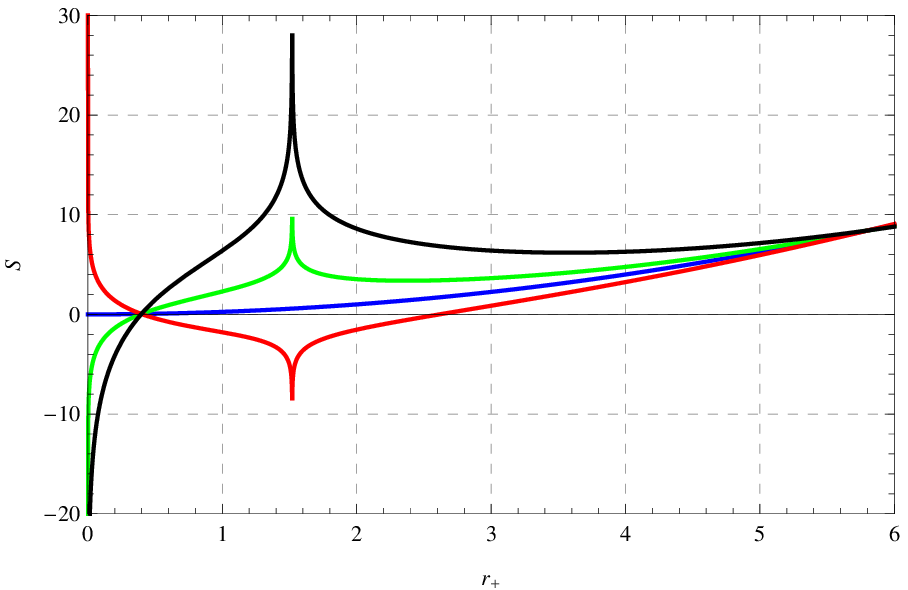} 
 \end{array}$
 \end{center}
\caption{Left: Entropy per volume vs. the black hole horizon radius for $d=3$, $l=1$ and $q=1$. Right:  Entropy per volume vs. the black hole horizon  radius for $d=3$, $l=2$ and $q=2$. Here, $\alpha=0$ denoted by blue line, $\alpha=-0.5$ denoted by green  line, $\alpha=0.5$ denoted by red line, and $\alpha=-1.5$ denoted by black line.}
 \label{fig1}
\end{figure}
 \begin{figure}[h!]
 \begin{center}$
 \begin{array}{cc}
\includegraphics[width=70 mm]{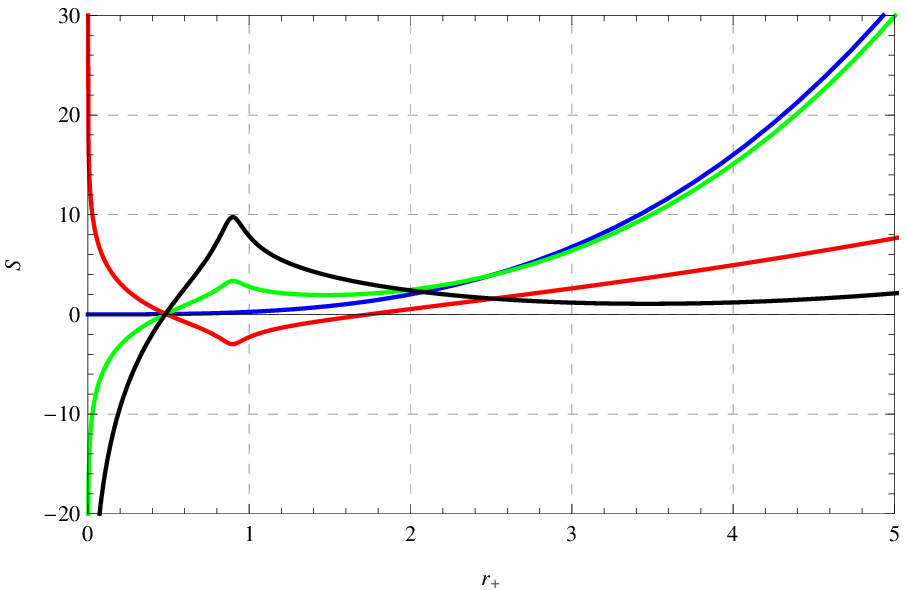}  \ \ \ \ & \includegraphics[width=70 mm]{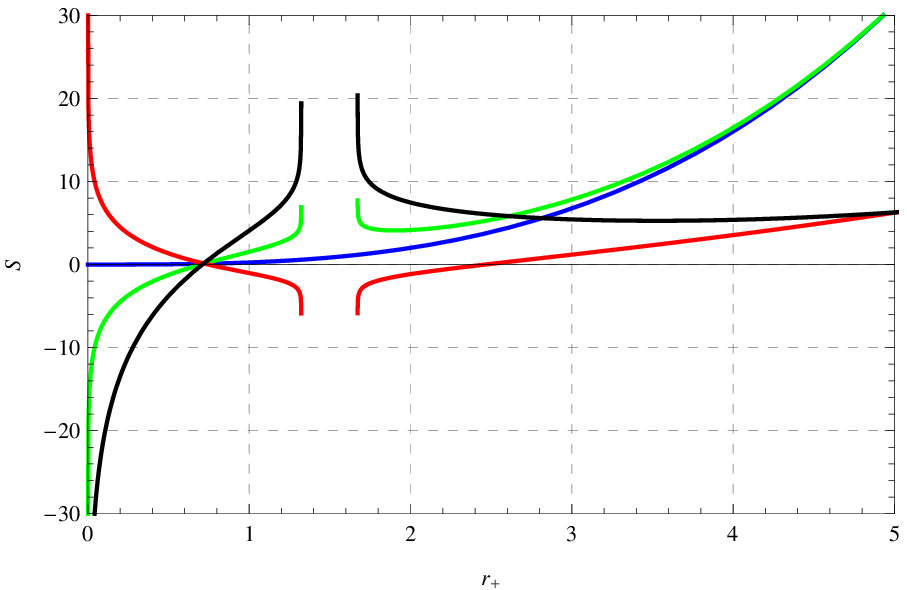} 
 \end{array}$
 \end{center}
\caption{Left: Entropy per volume vs. the black hole horizon radius for $d=4$, $l=1$ and $q=1$. Right:  Entropy per volume vs. the black hole horizon radius for $d=4$, $l=2$ and $q=2$. Here, $\alpha=0$ denoted by blue line, $\alpha=-0.5$ denoted by green  line, $\alpha=0.5$ denoted by red line, and $\alpha=-1.5$ denoted by black line.}
 \label{fig2}
\end{figure}
The effects of leading-order correction on  behavior of the entropy per volume with respect to horizon radius  can be seen in Figs. (\ref{fig1}) and (\ref{fig2}). For negative correction parameter $\alpha$, the first maxima (positive peak) occurs in between the critical points for the entropy per volume at sufficiently small
black holes.
Also, one can see in figures that in four space-time dimensions a negative region for the entropy density  occurs for quasitopological  black holes before second critical point corresponding to the positive values of correction parameter $\alpha$.
After the second critical point, the entropy density is an increasing function only.   For larger values of charge and AdS radius,   
  the critical value of entropy  density increases   and occurs at larger horizon radius. 
In five space-time dimensions case,  there exists only one critical point   and the first maxima/minima (peak) occurs after the critical point.  In this case, for larger values of charge and AdS radius, the corrected entropy density diverges  just after the critical point.  

The Gibbs free energy per unit volume for charged quasitopological black holes can be calculated utilizing the standard relation, $G(T_H,\Phi)=-\int SdT_H$, as follows
\begin{eqnarray}
G(T_H,\Phi) &=& -\frac{1}{16 \pi l^2}\left[r_+^d+\frac{d-2}{2(d-1)}l^2\Phi^2 r_+^{d-2}\right]+\alpha\frac{d(1+d)r_+}{4\pi l^2}
+\alpha\frac{ (d-3)(d-2)^2   }{8\pi (d-1)}\frac{\Phi^2}{r_+}\nonumber\\
 &-&\alpha\frac{ 2d(d-1)r_+^2-(d-2)^2l^2\Phi^2}{8\pi (d-1)l^2 r_+} \log\left[\frac{1}{64\pi^2}\left(\frac{d^2r^{d+1}}{l^4}+\frac{(d-2)^4}{4(d-1)^2}\Phi^4r_+^{d-3}\right.\right.\nonumber\\
 &-&\left.\left. \frac{d(d-2)^2}{(d-1)}\frac{\Phi^2r_+^{d-1}}{l^2}\right) \right].
 \end{eqnarray}
 Here, it is evident that in the limit $\alpha\rightarrow 0$, this coincides with the
 original expression calculated in Ref. \cite{l1}. 
  \begin{figure}[h!]
 \begin{center}$
 \begin{array}{cc}
\includegraphics[width=70 mm]{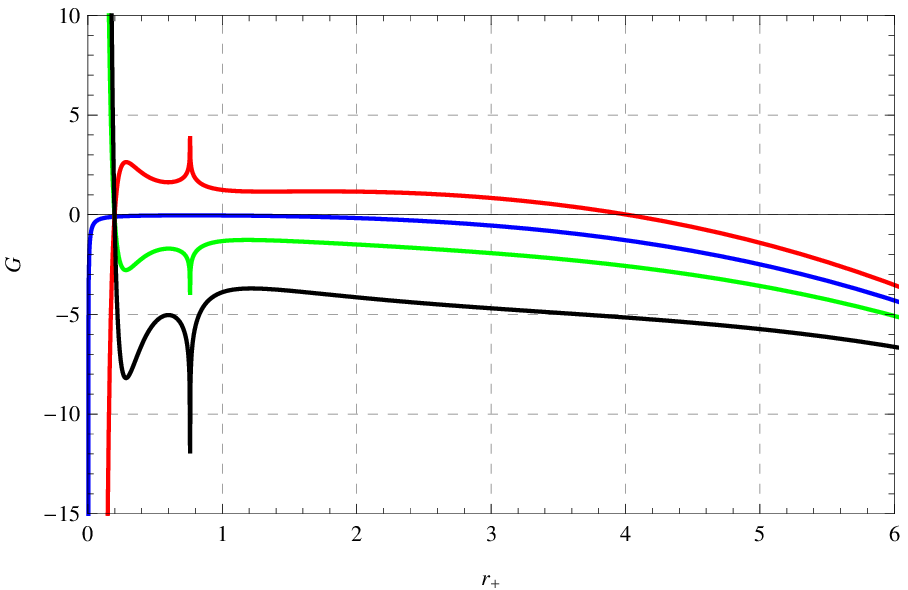}  \ \ \ \ & \includegraphics[width=70 mm]{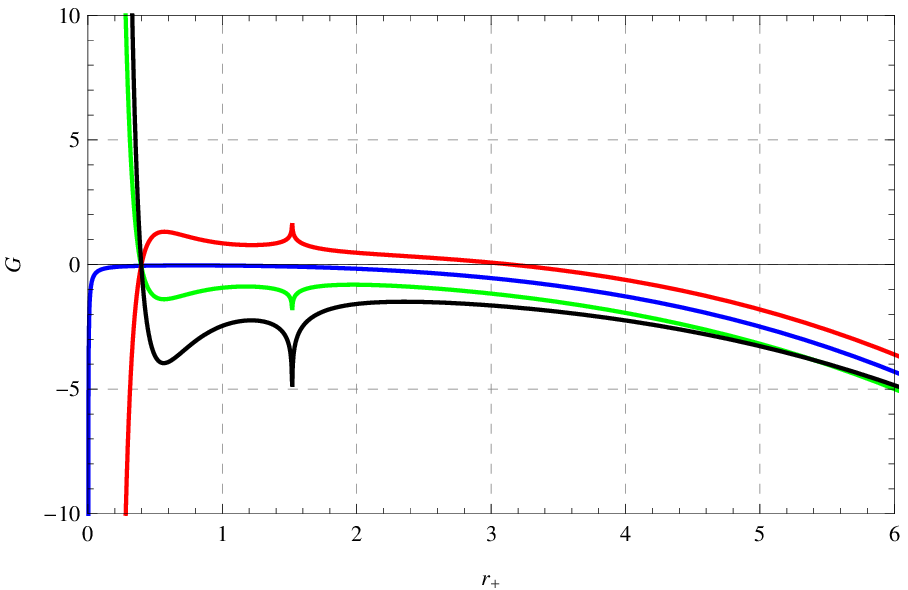} 
 \end{array}$
 \end{center}
\caption{Left: Gibbs free energy  per volume vs. the black hole horizon radius for $d=3$, $l=1$ and $q=1$. Right:  Gibbs free energy  per volume vs. the black hole horizon radius for $d=3$, $l=2$ and $q=2$. Here, $\alpha=0$ denoted by blue line, $\alpha=-0.5$ denoted by green  line, $\alpha=0.5$ denoted by red line, and $\alpha=-1.5$ denoted by black line.}
 \label{fig3}
\end{figure}
The effects of leading-order correction terms on the Gibbs free energy  per volume with respect to the black hole horizon radius in four space-time dimensions can be seen from Fig. (\ref{fig3}). We observe  that 
 the  Gibbs free energy  per volume is a decreasing function with respect to horizon radius. 
 The  Gibbs free energy  density
 without any correction is negligibly  small for smaller black holes and   becomes negative valued when   horizon radius increases. 
 However, the correction terms with negative correction parameter makes it finite negative valued for the smaller black holes. However,  the correction terms with positive correction parameter makes the Gibbs free energy density positive valued for the smaller black holes,  falls more sharply to take negative value along with increasing horizon radius. For  horizon radius $r_+\rightarrow 0$, asymptotic
  behavior of corrected  Gibbs free energy  per volume with negative $\alpha$ is completely opposite to 
 that of the uncorrected and corrected ones with positive $\alpha$.
    In fact, for sufficiently larger size of black hole the corrected  Gibbs free energy  per volume coincides the uncorrected one as expected.
 For the larger values of charge and AdS radius, the  corrected  Gibbs free energy  per volume behaves more closely to the uncorrected  one. 
  
The corrected charge density of charged quasitopological
black holes  under the influence of statistical fluctuations can be calculated 
as following: 
\begin{eqnarray}
Q&=&-\left(\frac{\partial G}{\partial\Phi}\right)_T,\nonumber\\
&=&\frac{1}{16\pi}\sqrt{ {2(d-1)(d-2)}}q +\frac{\alpha (d-2)^2}{4\pi}\sqrt{\frac{2(d-2)}{d-1}}\left[\frac{3d-(3d-1)q^2l^2r_+^{2-2d}}{d+(d-2)q^2l^2 r_+^{2-2d}} \right]\frac{q}{r^{d-1}}.
\end{eqnarray}
In limit $\alpha\rightarrow 0$, the above expression reduces to the original one  obtained in \cite{l1}. Here, we notice that for space  dimensions $d<3$, one can not have charged  quasitopological
black holes.   

The corrected expression for the mass per volume of the charged quasitopological
black holes can be easily calculated from the definition, $M=G+TS+\Phi Q$,  as follows
\begin{eqnarray}
M 
&=& \frac{(d-1)}{16\pi}m+\alpha\frac{(d-2)}{4\pi}\left[\frac{d(7d-15)- (d-2)(5d-1)q^2l^2 r_+^{2-2d}}{d+(d-2)q^2l^2 r_+^{2-2d}} \right]\frac{q^2}{r_+^{2d-3}}\nonumber\\
&+&\alpha\frac{d(d+1)}{4\pi l^2}r_+.
\end{eqnarray}
 \begin{figure}[h!]
 \begin{center}$
 \begin{array}{cc}
\includegraphics[width=70 mm]{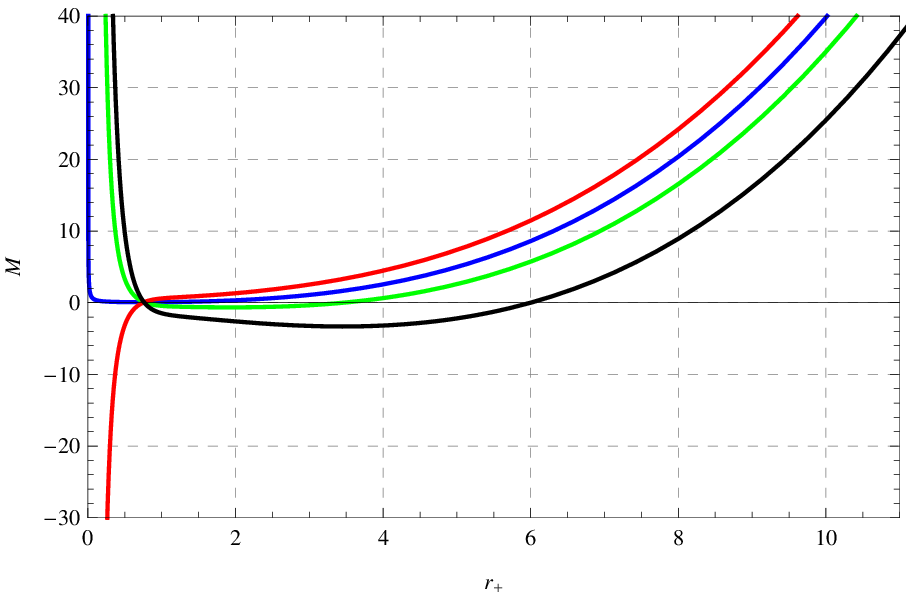}  \ \ \ \ & \includegraphics[width=70 mm]{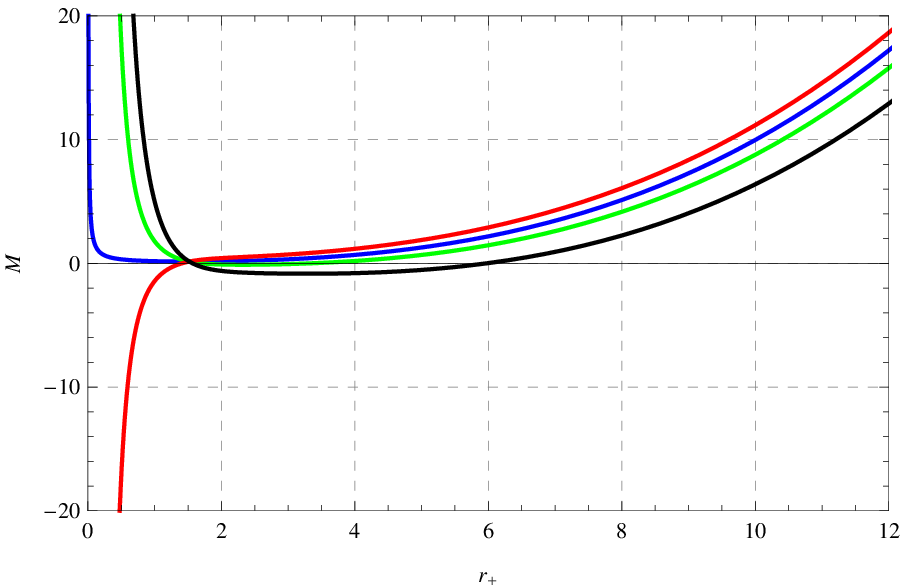} 
 \end{array}$
 \end{center}
\caption{Left: Mass  per volume vs. the black hole horizon radius for $d=3$, $l=1$ and $q=1$. Right:  Mass  per volume vs. the black hole horizon radius for $d=3$, $l=2$ and $q=2$. Here, $\alpha=0$ denoted by blue line, $\alpha=-0.5$ denoted by green  line, $\alpha=0.5$ denoted by red line, and $\alpha=-1.5$ denoted by black line.}
 \label{fig4}
\end{figure}
 A comparative analysis of corrected and uncorrected mass per volume can be seen in
 Fig. (\ref{fig4}). One can see, for sufficiently large  size of black holes, the corrected and uncorrected mass per volume show 
same behavior as expected. However, when horizon radius tends to zero value,
 the corrected mass per volume with positive correction parameter shows opposite behavior and
 takes negative asymptotic value. The larger values of charge and AdS radius minimize the differences of
 the  corrected  and uncorrected mass per volume.
 We note that   a critical value exists for the mass per volume for  small black holes
 after that the mass density becomes an increasing function. 
 \subsection{Stability of charged quasitopological black holes}
Now, we discuss thermal stability  of the charged quasitopological black holes.
It is well-known that the stability
conditions in canonical ensemble depend on sign of the specific heat. 
A change of sign may appear whether when specific heat meets root(s) or divergence(es). 
The root of specific capacity (or temperature) confirms a
bound point. This bound  point  divides physical solutions (which corresponds to positive temperature) from non-physical solutions (which corresponds to negative temperature).
However,  the divergences of specific heat    represent to the phase transition points. The
negative specific heat   represents to the unstable  solutions which may encounter a phase transition to acquire a stable state.  
 
The specific heat per volume with a fixed chemical potential ($\Phi$) is given by
\begin{eqnarray}
C_\Phi &=&T\left(\frac{\partial S}{\partial T}\right)_\Phi,\nonumber\\
&=&  \frac{2\pi (d-1)^2l^2 r_+^d}{2d(d-1)r_+^2+(d-2)^2l^2\Phi^2}  +2\alpha,\nonumber\\ 
&=&  \frac{\pi (d-1)l^2 r_+^{d-1}}{dr_+^{2d-2}+(d-2)l^2q^2 r_+^{3-2d}}   +2\alpha. 
\end{eqnarray}
In order to get   bound points, we solve the denominator
 of above expression with respect to horizon radius    and get
\begin{eqnarray}
r_c&=&\left[-\frac{(d-2)l^2q^2}{d} \right]^{1/4d-3}.
\end{eqnarray}
However, to get phase transition points, one can solve the numerator
 of above expression with respect to horizon radius which seems a cumbersome task for an arbitrary space-time dimensions  
 in   presence of
 correction parameter. 
 
 \begin{figure}[h!]
 \begin{center}$
 \begin{array}{cc}
\includegraphics[width=70 mm]{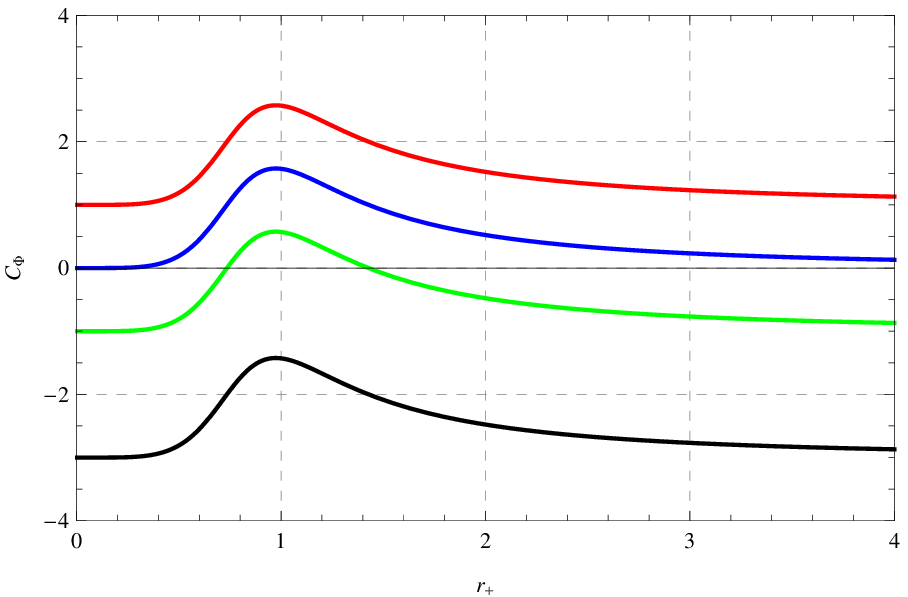}  \ \ \ \ & \includegraphics[width=70 mm]{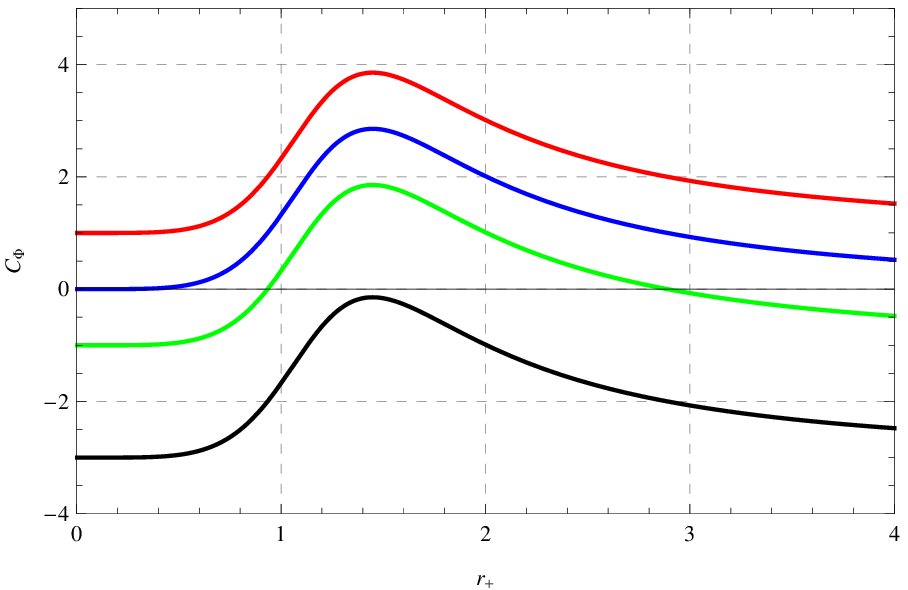} 
 \end{array}$
 \end{center}
\caption{Left: Specific heat per volume vs. the black hole horizon radius for $d=3$, $l=1$ and $q=1$. Right:  Specific heat vs. the black hole horizon radius for $d=3$, $l=2$ and $q=2$. Here, $\alpha=0$ denoted by blue line, $\alpha=-0.5$ denoted by green  line, $\alpha=0.5$ denoted by red line, and $\alpha=-1.5$ denoted by black line.}
 \label{fig5}
\end{figure}
From the Fig. (\ref{fig5}), we observe that for  quasitopological black holes 
in four space-time dimensions there exists no phase transition point  corresponding to both the
uncorrected and corrected specific heat per volume with positive $\alpha$ and black  holes are 
 stable.  
Interestingly, we find that  the correction term
with negative correction parameter causes instability to the black holes and
 a stable state  exists
only   in case of corrected specific heat density with smaller value of negative correction parameter.

\section{Charged rotating quasitopological black holes: Thermal instability} \label{4} 
In order to describe the charged rotating quasitopological black holes,   we 
 equip our charged static solution with a global rotation. The metric for a (d+1)-dimensional asymptotically
AdS rotating solution with $k$ rotation parameters can
be written as \cite{l1}
\begin{eqnarray}
ds^2&=&N^2(r)f(r)\left(\Delta dt -\sum_{i=1}^{k}a_id\phi_i\right)^2 +\frac{r^2}{l^4}\sum_{i=1}^{k}(a_idt-\Delta l^2d\phi_i)^2+\frac{dr^2}{f(r)}\nonumber\\
&-&\frac{r^2}{l^2}\sum_{i< j}^{k}(a_id\phi_j - a_jd\phi_i)^2+r^2\sum_{i=k+1}^{d-1}d\phi_i^2, 
\end{eqnarray}
 where $\Delta^2 =1+\sum_{i=1}^k\frac{a_i^2}{l^2}$ and 
 the angular coordinates can have following range: $0\leq \phi_i<2\pi$.
Also, the gauge potential corresponding to this metric has following form: $A_a(r)=-\sqrt{\frac{2(n-1)}{ n-2 }}\frac{q}{r^{d-2}}(\Delta dt -\Sigma_{i=1}^k a_i d\phi_i)$.

 The Hawking temperature from the area law is calculated by
 \begin{eqnarray}
T_H=\left.\frac{f'(r)}{4\pi\Delta}\right|_{r=r_+}=\frac{dr_+-(d-2)q^2l^2r^{3-2d}}{4\pi \Delta 
l^2}.\label{haw}
\end{eqnarray}
 The horizon radius in terms of the intensive quantities can be written as 
\begin{eqnarray}
r_+=(1-l^2\Omega^2)^{-1/2}\left\{\frac{2\pi l^2}{d}T+2\left[\frac{\pi^2 l^4}{d^2}T^2+\frac{(d-2)^2 l^2}{8d(d-1)}\Phi^2
\right]^{1/2}\right\},
\end{eqnarray}
where $\Phi$ is the electric potential, measured at infinity with respect
to the horizon, with following explicit form \cite{l1}:
\begin{eqnarray}
\Phi=\sqrt{\frac{2(d-1)}{d-2}}\frac{q}{\Delta r_+^{d-2}}.
\end{eqnarray}
The entropy density of charged rotating quasitopological black hole without any thermal fluctuation can be calculated
with the help of Gibbs free energy function  
and the temperature  as \cite{l1} 
\begin{eqnarray}
S_0=\frac{\Delta}{4}r_+^{d-1}.\label{ent}
\end{eqnarray} 
Due to thermal fluctuation around equilibrium induces a correction to the 
original entropy density. We calculate this first-order corrected entropy density as
\begin{eqnarray}
S =\frac{\Delta}{4}r_+^{d-1}+\alpha \log\left[\frac{d^2r_+^{d+1}+(d-2)^2q^4l^4 r_+^{5-3d}-2d(d-2)q^2l^2r_+^{3-d}}{64\pi^2 \Delta l^4}\right],
\end{eqnarray}
where   relations (\ref{correctedS}), (\ref{haw}) and (\ref{ent}) have been utilized.
\begin{figure}[h!]
 \begin{center}$
 \begin{array}{cc}
\includegraphics[width=70 mm]{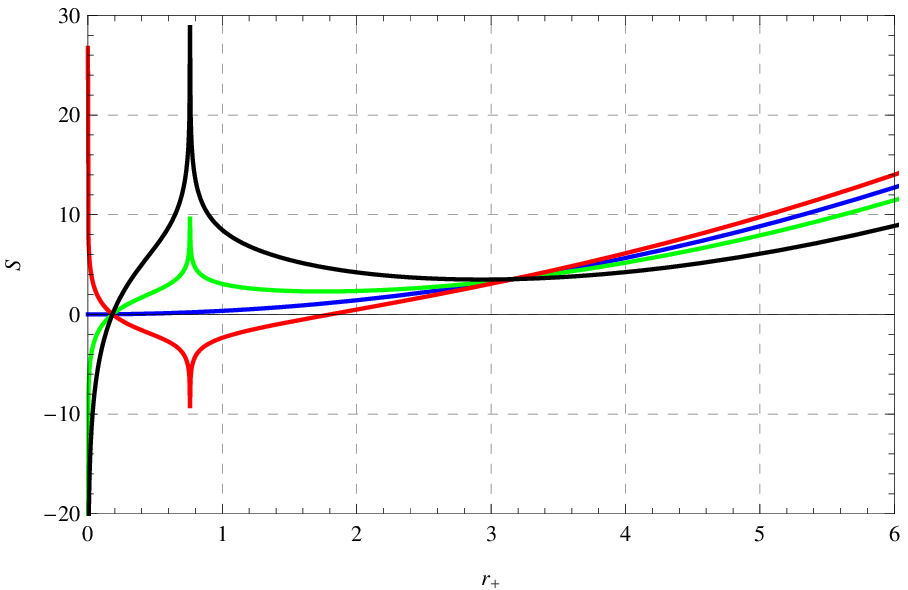}  \ \ \ \ & \includegraphics[width=70 mm]{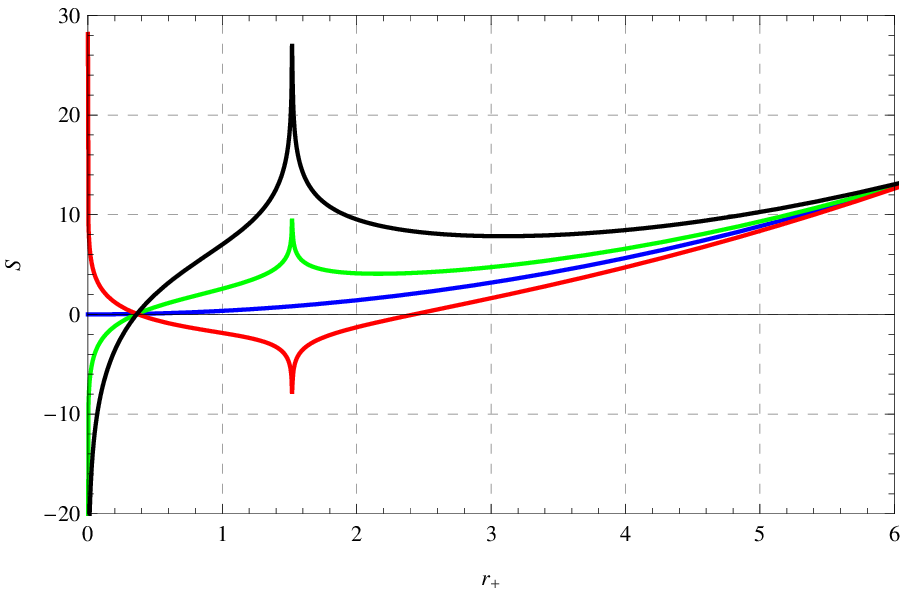} 
 \end{array}$
 \end{center}
\caption{Left: Entropy per volume vs. the black hole horizon radius for  $\Delta^2=2$, $d=3$, $l=1$ and $q=1$. Right:  Entropy per volume vs. the black hole horizon radius for  $\Delta^2=2$, $d=3$, $l=2$ and $q=2$. Here, $\alpha=0$ denoted by blue line, $\alpha=-0.5$ denoted by green  line, $\alpha=0.5$ denoted by red line, and $\alpha=-1.5$ denoted by black line.}
 \label{fig6}
\end{figure}
 \begin{figure}[h!]
 \begin{center}$
 \begin{array}{cc}
\includegraphics[width=70 mm]{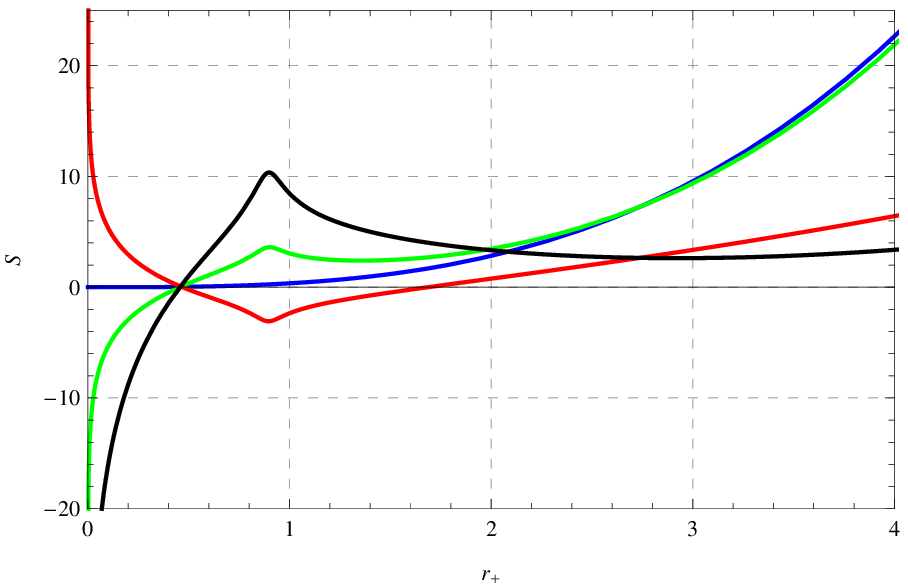}  \ \ \ \ & \includegraphics[width=70 mm]{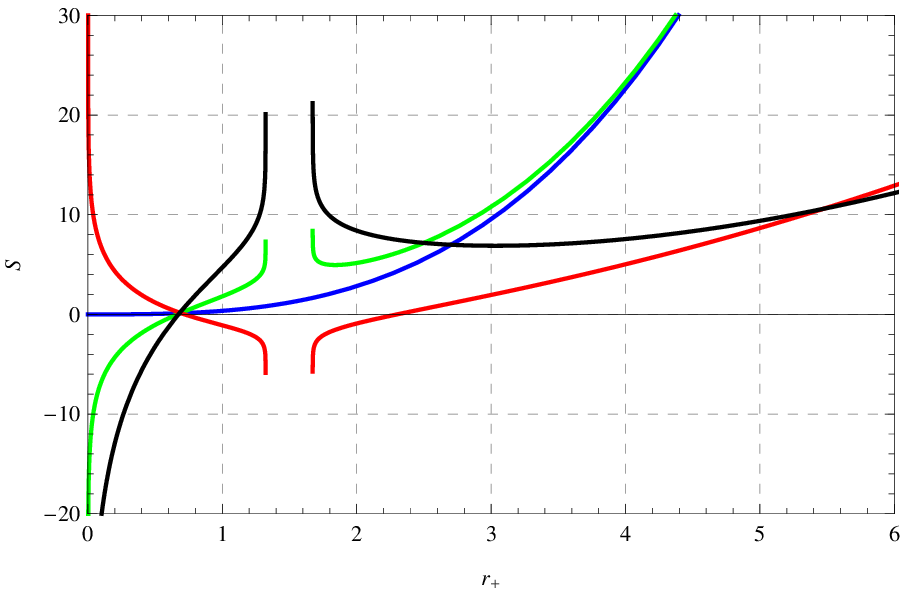} 
 \end{array}$
 \end{center}
\caption{Left: Entropy per volume vs. the black hole horizon radius for $\Delta^2=2$, $d=4$, $l=1$ and $q=1$. Right:  Entropy per volume vs. the black hole horizon radius for $\Delta^2=2$, $d=4$, $l=2$ and $q=2$. Here, $\alpha=0$ denoted by blue line, $\alpha=-0.5$ denoted by green  line, $\alpha=0.5$ denoted by red line, and $\alpha=-1.5$ denoted by black line.}
 \label{fig7}
\end{figure}
A comparative study of  leading-order corrected and uncorrected entropy  densities with respect to horizon radius  for four and five space-time dimensions can be seen in Figs. (\ref{fig6}) and (\ref{fig7}) respectively. For negative correction parameter $\alpha$ there exists first maxima  (positive peak) for the entropy per volume in between the critical points. However, there exists a negative region for entropy density with a minima (negative peak) corresponding to positive $\alpha$ which is physically irrelevant and can be ignored. 
One can see that  the correction terms do not play an important role 
   for the entropy per volume at  sufficiently larger horizon radius. Also, there exist  critical   entropy densities at horizon radii $r_+=\approx 0.2$ and $r_+=\approx 3$  for four dimensional black holes. For larger values of charge and AdS radius,   
  the second  critical value of entropy density  increases and occurs at larger horizon radius. 
  For five space-time dimensions case,  there exists only one critical point fro entropy density  and the corrected entropy density with large negative parameter falls more sharply.  As   the charge and AdS radius take larger values, the corrected entropy density diverges after the first critical point.

The corrected expression for the Gibbs free energy per unit volume is calculated by
\begin{eqnarray}
G(T_H,\Phi,\Omega)  
&=& -\frac{1}{16 \pi l^2}\left[r_+^d+\frac{d-2}{2(d-1)(1-l^2\Omega^2)}l^2\Phi^2 r_+^{d-2}\right]+\alpha\frac{d(1+d)r_+}{4\pi\Delta l^2}
\nonumber\\
 &+&\alpha\frac{ (d-3)(d-2)^2  }{8\pi (1-d)\Delta }\frac{\Phi^2}{r_+}-\alpha\frac{ 2d(d-1)r_+^2-(d-2)^2l^2\Phi^2}{8\pi (1-d)\Delta l^2 r_+} \times\nonumber\\
 &&\log\left[\frac{1}{64\pi^2}\left(\frac{d^2r^{d+1}}{l^4}+\frac{(d-2)^4}{4(d-1)^2}\Phi^4r_+^{d-3}- \frac{d(d-2)^2}{(d-1)}\frac{\Phi^2r_+^{d-1}}{l^2}\right) \right]\nonumber\\
 & +&\alpha \frac{ \log [\Delta] }{ \Delta}T_H. 
 \end{eqnarray}
where $\Omega$ is the angular velocity of the Killing horizon and has following form: $\Omega_i=a_i/\Delta l^2$.
 \begin{figure}[h!]
 \begin{center}$
 \begin{array}{cc}
\includegraphics[width=70 mm]{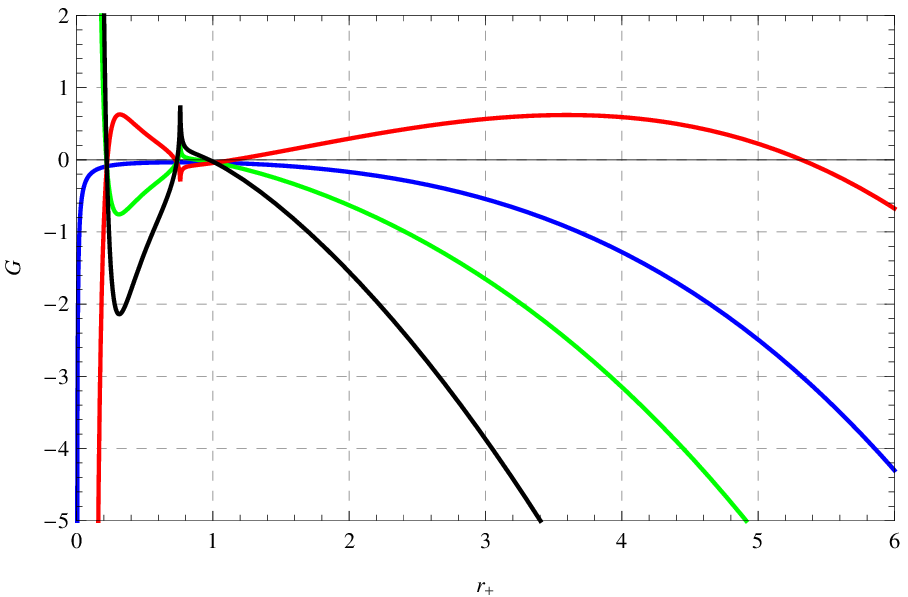}  \ \ \ \ & \includegraphics[width=70 mm]{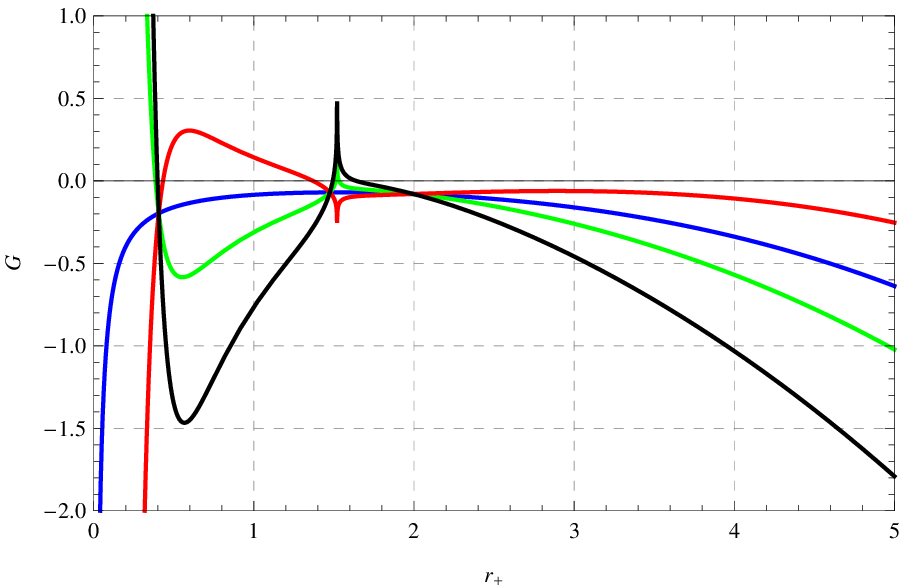} 
 \end{array}$
 \end{center}
\caption{Left: Gibbs free energy  per volume vs. the black hole horizon radius for $d=3$, $\Delta^2=2$, $l=1$ and $q=1$. Right:  Gibbs free energy  per volume vs. the black hole horizon radius for $d=3$, $\Delta^2=2$, $l=2$ and $q=2$. Here, $\alpha=0$ denoted by blue line, $\alpha=-0.5$ denoted by green  line, $\alpha=0.5$ denoted by red line, and $\alpha=-1.5$ denoted by black line.}
 \label{fig8}
\end{figure}
We draw a plot in Fig. (\ref{fig8}) for the Gibbs
free energy density with respect to horizon radius to make a comparative discussion between the corrected and uncorrected  the Gibbs
free energy densities. In this figure, we see that  the   Gibbs free energy density is a negative valued function   for the larger horizon radius. The leading-order correction terms with (negative-)positive  $\alpha$ make  it (more-)less negative valued for larger horizon radius. For small horizon radius, the corrected Gibbs energy density   with positive $\alpha$ is  positive valued. In the limit $r_+\rightarrow 0$, the corrected
Gibbs free energy density with negative $\alpha$ shows opposite asymptotic behavior in comparison to uncorrected and corrected ones with positive $\alpha$. Two critical points occur for the Gibbs free energy density. 
    For the larger values of charge and AdS radius, after critical points
the corrected Gibbs free energy per volume with negative $\alpha$ becomes less negative. 

Now, we calculate the    charge  density  of charged rotating quasitopological 
black hole  under the influence of statistical fluctuations. This is given by,  
\begin{eqnarray}
Q&=&-\left(\frac{\partial G}{\partial\Phi}\right)_T,\nonumber\\
&=&\frac{1}{16\pi}\sqrt{ {2(d-1)(d-2)}}\Delta q +\frac{\alpha (d-2)^2}{4\pi\Delta}\sqrt{\frac{2(d-2)}{d-1}}\left[\frac{3d-(3d-1)q^2l^2r_+^{2-2d}}{d+(d-2)q^2l^2 r_+^{2-2d}} \right]\frac{q}{r^{d-1}}.
\end{eqnarray}
From the above expression, the original expression of total    charge  density  given in 
\cite{l1} can be recovered in the   $\alpha\rightarrow 0$  limit. Also, we see that
the charged rotating quasitopological black hole does not exist for space dimensions $d<3$.

Since the black hole solution is endowed with a global rotation, therefore, this possesses an associated   angular momentum also. 
We compute the first-order corrected angular momentum per volume  as follows,
\begin{eqnarray}
J_i&=&-\left(\frac{\partial G}{\partial \Omega_i}\right)_{T,\Phi},\nonumber\\
&=&\frac{d}{16\pi}\Delta m a_i-\frac{\alpha}{2\pi}\left[\frac{d^2 r_+}{l^2}-(d-2)\frac{q^2}{r_+^{2d-3}} \right]a_i\nonumber\\
&+&\frac{\alpha}{4\pi}\left[\frac{d  r_+}{l^2}+\frac{(d-2)q^2}{r_+^{2d-3}} \right]a_i
\log\left[\frac{d^2r_+^{d+1}+(d-2)^2q^4l^4 r_+^{5-3d}-2d(d-2)q^2l^2r_+^{3-d}}{64\pi^2 \Delta l^4}\right].
\end{eqnarray}
This expression  also coincides with the original one calculated in \cite{l1},
when we switch off thermal fluctuations (i.e. $\alpha$ =0).  

Now, utilizing the standard relation, $M=G+TS+\Phi Q+\sum_{i=1}^k \Omega_i J_i$, 
 we are able to calculate the first-order corrected total mass per volume of the charged rotating quasitopological 
black hole   as following:
\begin{eqnarray}
M
&=&  \frac{1}{16\pi}(d\Delta^2-1) m+\frac{\alpha}{4\pi\Delta l^2}d(3d-2d\Delta^2 +1)r_+  +\alpha\frac{\log[\Delta]}{\Delta} T_H\nonumber\\
&-&\frac{\alpha(d-2)}{4\pi\Delta}\left[\frac{1-2\Delta^2(\Delta^2-1)}{\Delta^2}-2(d-2)\left(\frac{3d-(3d-1)q^2l^2r_+^{2-2d}}{d+(d-2)q^2l^2 r_+^{2-2d}} \right) \right]\frac{q^2}{r_+^{2d-3}}\nonumber\\
&+&\alpha\frac{(\Delta^2+1)d r_++(\Delta^2 -1)\Delta^{-2}(d-2)q^2l^2r_+^{3-2d} }{4\pi \Delta l^2} \times\nonumber\\
&&\log\left[\frac{d^2r_+^{d+1}+(d-2)^2q^4l^4 r_+^{5-3d}-2d(d-2)q^2l^2r_+^{3-d}}{64\pi^2 \Delta l^4}\right].
\end{eqnarray}
This expression of corrected total mass  density is also consistent with the one calculated originally in \cite{l1} in the limit $\alpha\rightarrow 0$.

 \begin{figure}[h!]
 \begin{center}$
 \begin{array}{cc}
\includegraphics[width=70 mm]{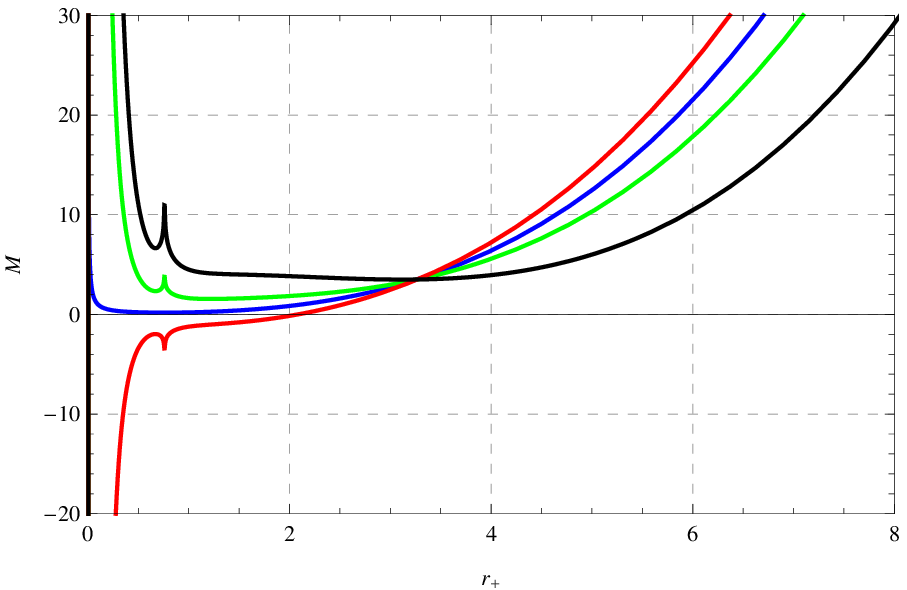}  \ \ \ \ & \includegraphics[width=70 mm]{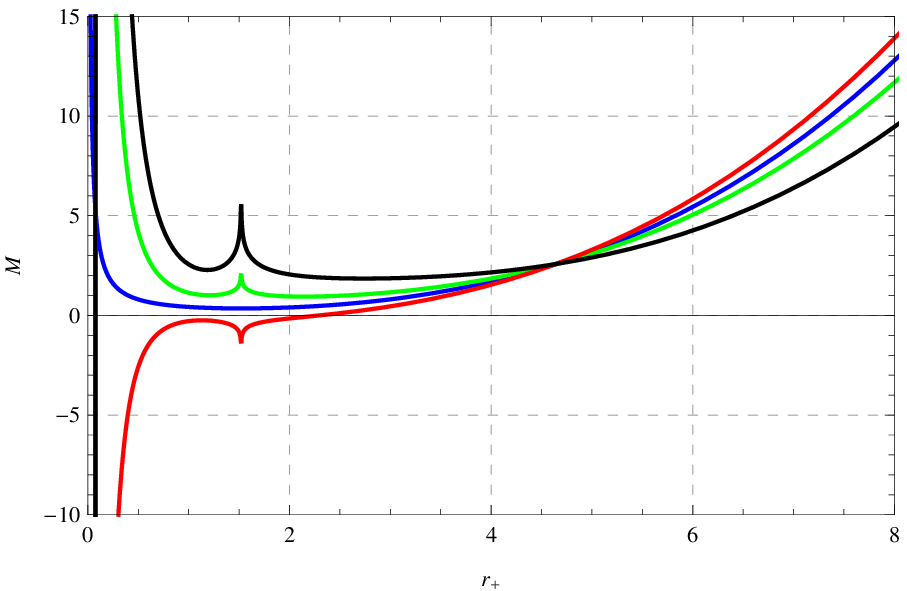} 
 \end{array}$
 \end{center}
\caption{Left: Mass  per volume vs. the black hole horizon radius for $d=3$, $\Delta^2=2$, $l=1$ and $q=1$. Right:  Mass  per volume vs. the black hole horizon radius for $d=3$, $\Delta^2=2$, $l=2$ and $q=2$. Here, $\alpha=0$ denoted by blue line, $\alpha=-0.5$ denoted by green  line, $\alpha=0.5$ denoted by red line, and $\alpha=-1.5$ denoted by black line.}
 \label{fig9}
\end{figure}
In order to discuss the effects of thermal fluctuations on the total mass density as one reduces the size of the charged rotating quasitopological black hole, we plot Fig. (\ref{fig9}). 
We see that the corrected total mass density 
with negative $\alpha$ is an decreasing  function before the critical point and  increasing function  after the critical point, but  not a negative valued function.
 However, the corrected   mass density  with positive $\alpha$ is an increasing function only but  a negative valued function before the critical point.
The larger values of charge and AdS radius  decrease the critical value of mass density a bit and  occurs
at bit larger horizon radius. 
 
 \subsection{Stability of charged rotating quasitopological black holes}
Here, in order to discuss the thermal stability   for charged rotating quasitopological black holes, we would analyse the sign of the specific heat as the negative specific heat represents to the unstable solutions which may
encounter a phase transition to acquire a stable state.
 
The specific heat per volume with a fixed chemical potential ($\Phi$) for the charged rotating quasitopological black holes is calculated by
\begin{eqnarray}
C_\Phi &=&T\left(\frac{\partial S}{\partial T}\right)_\Phi,\nonumber\\
&=&  \frac{2\pi (d-1)^2l^2 \Delta^2 r_+^d}{2d(d-1)r_+^2+(d-2)^2l^2\Phi^2}  +2\alpha,\nonumber\\ 
&=&  \frac{\pi (d-1)l^2  \Delta^2 r_+^{d-1}}{dr_+^{2d-2}+(d-2)l^2q^2 r_+^{3-2d}}   +2\alpha. 
\end{eqnarray}
The bound points can be obtained by solving  the denominator
 of above expression with respect to horizon radius.  By doing so, we  obtain
\begin{eqnarray}
r_c&=&\left[-\frac{(d-2)l^2q^2}{d} \right]^{1/4d-3},
\end{eqnarray}
which is  an exactly same point as obtained in the case of charged quasitopological black holes without rotation. 

  \begin{figure}[h!]
 \begin{center}$
 \begin{array}{cc}
\includegraphics[width=70 mm]{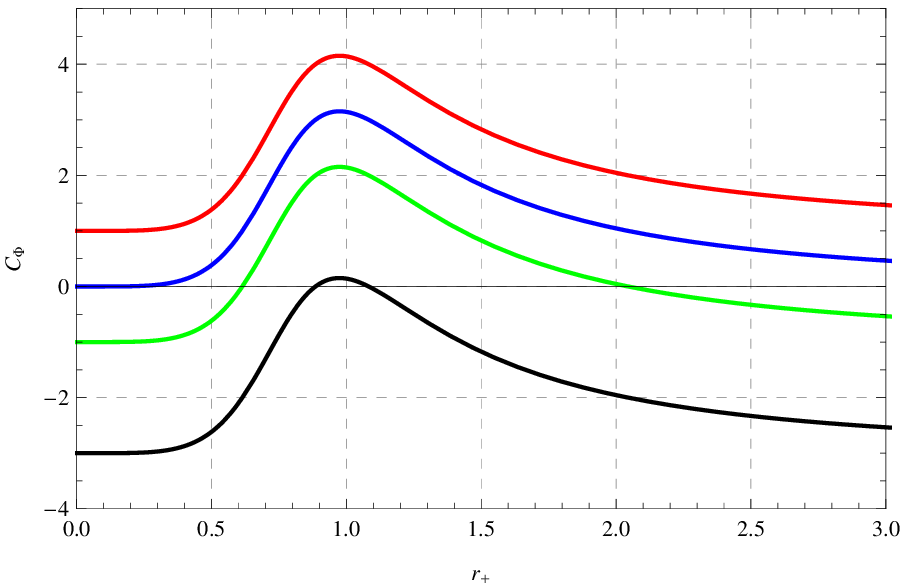}  \ \ \ \ & \includegraphics[width=70 mm]{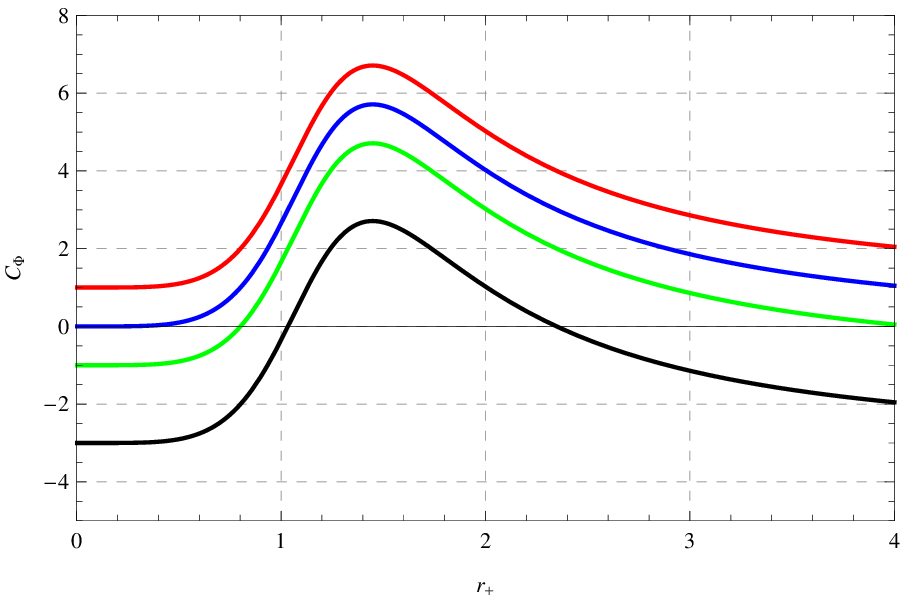} 
 \end{array}$
 \end{center}
\caption{Left: Specific heat per volume  vs. the black hole horizon radius for $\Delta^2=2$, $d=3$, $l=1$ and $q=1$. Right:  Specific heat vs. the black hole horizon radius for  $\Delta^2=2$, $d=3$, $l=2$ and $q=2$. Here, $\alpha=0$ denoted by blue line, $\alpha=-0.5$ denoted by green  line, $\alpha=0.5$ denoted by red line, and $\alpha=-1.5$ denoted by black line.}
 \label{fig10}
\end{figure}
In order to see the effects of thermal fluctuations on the stability of charged rotating quasitopological black hole, we plot the   Fig. (\ref{fig10}). We see that due to the
correction terms with negative correction parameters instabilities occur for the charged rotating quasitopological black holes. However, the  
correction terms with positive correction parameter make the specific heat more positive valued and therefore   more stable. 
The larger values of charge and AdS radius improves the stability of such black holes 
corresponding to the correction terms with  negative correction parameters.

\section{Concluding remarks}\label{5}
It is well-known that quasitopological gravity is a new gravitational theory, including Gauss-
Bonnet term and curvature-cubed 
interactions,  which  possesses exact black hole solutions. Here, we have considered both the charged
and charged rotating quasitopological gravity with black hole solutions to study the 
effects of thermal fluctuation on thermodynamics of small black holes.

First, we have evaluated the  leading-order correction to the 
entropy density of charged quasitopological black hole and made 
a comparative analysis  between corrected and  uncorrected entropy densities through plots for small
 sizes of the black holes. We have  found that corresponding to  (negative-)positive correction parameters  there exist 
  (positive-)negative peaks for the corrected entropy density in between the critical points.  Also, corrected  entropy density   becomes negative valued
  corresponding to the positive values of correction parameter, which is not physical and therefore can be forbidden.    The correction term plays a crucial role for  entropy densities in between these critical points. Furthermore, we have computed the first-order corrected Gibbs free energy density. We have plotted graph to make the comparative analysis and 
  found that the leading-order corrected Gibbs free
energy density   with negative correction parameter makes it (more-)less  negative valued for the (smaller-)larger  black holes.
In spite of that  the corrected  Gibbs free energy density with positive correction parameter  became more positive valued.  The higher values of charge and AdS radius decrease  the deviation  of corrected Gibbs free energy density  to that of the uncorrected one. We have also calculated the corrected expression for total charge of
quasitopological black holes. Finally, we have evaluated the more exact expression for  the
total mass density of such black holes. A critical horizon radius has been found for total mass density and before which the positive correction parameter causes opposite behavior. 
We have also discussed the stability of such black holes by calculating the corrected specific heat density with
fixed chemical potential  and have found a bound point. We noticed that a phase transition does not exist for quasitopological black holes under the influence of thermal fluctuation with positive $\alpha$ and therefore black holes are in stable state. 
However, due to the thermal fluctuation with negative  $\alpha$  an instability occurs
 such black holes.  

Furthermore, we have considered a charged quasitopological black holes endowed with the
global rotation and have computed the Hawking temperature and leading-order correction to the 
entropy density. We have found that a maxima (positive peak) occurs for the corrected entropy density with negative  $\alpha$   in between the critical horizon radii. However, 
the corrected entropy density becomes negative for positive $\alpha$
which is physically irrelevant and can be ignored. This indicates that only negative valued 
correction parameter $\alpha$ is physically relevant. 
 We also noted that for  larger values of charge and AdS radius  the second critical   point occurs at larger horizon radius.  We have obtained the corrected expressions for Gibbs free energy, 
 charge, angular momentum and total mass densities.  The correction terms with  
 (negative-)positive  $\alpha$ make the Gibbs free energy density (more-)less negative valued.
 The  corrected Gibbs free energy density  with negative $\alpha$ shows opposite 
 asymptotic behavior.   The larger values of charge and AdS radius make Gibbs free energy per volume   more negative valued.  We have noticed that the correction terms with negative $\alpha$   increase  total mass density   before   critical point and decrease   
    after critical point.  
 However, the corrected mass density with positive $\alpha$  takes negative asymptotic value
as horizon radius tends to zero. Also, the larger values of charge and AdS radius  increase the value of critical  horizon radius. We have calculated the corrected specific heat with fixed chemical potential in case of  charged rotating quasitopological black holes  also and discussed their stability. It would be interesting to investigate the effects of thermal fluctuation on the $P-V$
criticality of quasitopological black holes where negative cosmological constant could
play  the role of thermodynamic pressure.

\end{document}